\documentclass[11pt,journal,letterpaper,twocolumn,twoside,nofonttune]{IEEEtran}
\IEEEoverridecommandlockouts

\usepackage[utf8]{inputenc}
\usepackage[T1]{fontenc}
\usepackage{url}
\usepackage{ifthen}
\usepackage{cite}
\usepackage[cmex10]{amsmath}
\usepackage[shortlabels]{enumitem}
\usepackage[compact]{titlesec}
\titlespacing{\section}{0pt}{*0.9}{*0.5}
\titlespacing{\subsection}{0pt}{*0.8}{*0.4}

\usepackage[top=1in, bottom=1in, left=1.25in, right=1.25in]{geometry}

\usepackage{amsmath}
\usepackage{amsthm}
\usepackage{amsfonts}
\usepackage{amssymb}
\usepackage{mathtools}
\usepackage{color}
\usepackage{verbatim}
\usepackage{mathtools}
\usepackage{mathrsfs}
\usepackage[normalem]{ulem}
\usepackage{enumitem}
\usepackage{hyperref}
\usepackage[ruled,vlined]{algorithm2e}
\usepackage{algpseudocode}
\usepackage{balance}
\usepackage{graphicx}
\usepackage{floatrow}
\usepackage{cleveref}

\definecolor{purple}{rgb}{0.5, 0.0, 0.5}

\definecolor{mygray}{gray}{0.6}
\newcommand{\gray}{\color{mygray}}

\newcommand{\Bcal}{\mathcal{B}}

\newcommand{\I}{\mathcal{I}}
\newcommand{\J}{\mathcal{J}}

\newcommand{\ow}{\mathcal{O}}

\newcommand{\Vb}{\bold{V}}
\newcommand{\wb}{\bold{w}}
\newcommand{\Wb}{\bold{W}}
\newcommand{\xb}{\bold{x}}
\newcommand{\Xb}{\bold{X}}

\newcommand{\yb}{\bold{y}}

\newcommand{\Gb}{\bold{G}}
\newcommand{\Gbt}{\tilde{\Gb}}

\newcommand{\D}{\mathcal{D}}

\newcommand{\wbt}{\tilde{\bold{w}}}

\newcommand{\pt}{\tilde{p}}
\newcommand{\Scal}{\mathcal{S}}

\newcommand{\Z}{\mathbb{Z}}
\newcommand{\R}{\mathbb{R}}
\newcommand{\C}{\mathbb{C}}
\newcommand{\E}{\mathbb{E}}
\newcommand{\F}{\mathbb{F}}
\newcommand{\N}{\mathbb{N}}
\newcommand{\gb}{\bold{g}}
\newcommand{\pb}{\bold{p}}
\newcommand{\Pb}{\bold{P}}
\newcommand{\Sb}{\bold{S}}
\newcommand{\Sbt}{\tilde{\bold{S}}}

\newcommand{\Sbwt}{\widetilde{\bold{S}}}
\newcommand{\Ab}{\bold{A}}

\newcommand{\Abt}{\tilde{\Ab}}
\newcommand{\Abwh}{\widehat{\Ab}}
\newcommand{\Bb}{\bold{B}}

\newcommand{\Bbt}{\tilde{\Bb}}
\newcommand{\Bbwh}{\widehat{\Bb}}
\newcommand{\Cb}{\bold{C}}

\newcommand{\Cbwh}{\widehat{\Cb}}

\newcommand{\Db}{\bold{D}}

\newcommand{\Hb}{\bold{H}}
\newcommand{\Hbh}{\hat{\Hb}}
\newcommand{\Mb}{\bold{M}}

\newcommand{\Ub}{\bold{U}}
\newcommand{\Ubwh}{\widehat{\Ub}}

\newcommand{\Ib}{\bold{I}}
\newcommand{\Zb}{\bold{Z}}

\newcommand{\ab}{\bold{a}}

\newcommand{\bb}{\bold{b}}
\newcommand{\bbt}{\tilde{\bb}}

\newcommand{\Psit}{\tilde{\Psi}}
\newcommand{\Phib}{\bold{\Phi}}

\newcommand{\Omb}{\bold{\Omega}}
\newcommand{\Pib}{\bold{\Pi}}
\newcommand{\Sigb}{\bold{\Sigma}}
\newcommand{\mub}{\bold{\mu}}
\newcommand{\dt}{\tilde{d}}
\newcommand{\qt}{\tilde{q}}
\newcommand{\rpm}{\raisebox{.2ex}{$\scriptstyle\pm$}}

\newcommand{\sfsty}[1]{\ensuremath{\mathsf{#1}}}  

\newcommand{\rk}{\mathrm{rank}}

\newcommand*{\herm}{\mathsf{H}}
\newcommand{\bmm}{\sfsty{BasicMatrixMultiplication}}

\DeclareMathAlphabet{\mathpzc}{OT1}{pzc}{m}{it}
\newcommand{\brs}{\sfsty{BRS}}
\newcommand{\x}{\sfsty{x}}
\newcommand{\y}{\sfsty{y}}
\newcommand{\z}{\sfsty{z}}

\DeclarePairedDelimiter\floor{\lfloor}{\rfloor}

\newtheorem{Thm}{Theorem}

\begin{document}
\title{\huge Approximate Distributed Coded Computing:\\ Polynomial Codes and Randomized Sketching}

\author{
  \IEEEauthorblockN{\small\textbf{Neophytos Charalambides} (ncharalambides@ucsd.edu), \textbf{Arya Mazumdar} (arya@ucsd.edu)}
  \\ \IEEEauthorblockA{{\small Department of CSE and Halicio\u glu Data Science Institute, University of California, San Diego}
  }
\vspace{-10mm}
}

\maketitle

\begin{abstract}
Coded computing is a distributed paradigm that uses coding theory to introduce \textit{redundancy} and overcome bottlenecks in large-scale systems. In the same vein, randomized numerical linear algebra employs probabilistic methods to \textit{compress} and accelerate linear algebraic operations, addressing challenges in high-dimensional data analysis. This article reviews the foundations of both fields and presents distributed schemes that combine techniques from both to speed up optimization and machine learning algorithms, in the presence of slow or non-responsive servers. Along the way, we touch on various related topics and mathematical concepts.
\end{abstract}

\vspace{-1mm}
\section{Introduction}
\label{intro}

The advent of massive datasets has made distributed computing essential for modern data-intensive applications, enabling mathematical programming and large-scale machine learning to scale with data size. Consequently, such systems often suffer from slow or failing servers, known as \textit{stragglers}. To alleviate latencies, coding-theoretic techniques have emerged as tools for recovering exact or approximate results in the presence of stragglers. On the other hand, speeding up computations via randomized dimensionality reduction has been a promising alternative for efficient approximations over the past two decades.

These two paradigms, \textit{coded computing} (CC)~\cite{LLPPR17,LA20} and \textit{randomized numerical linear algebra} (RandNLA or ``Sketching'')~\cite{Wan15,murray2023}, tackle large-scale problems in complementary ways. CC introduces \textit{redundancy} through coding theory, while sketching induces \textit{compression}, leveraging randomness and sampling. In information-theoretic terms, they mirror channel and source coding, and at a high level, CC and sketching address complementary bottlenecks: \textit{reliability} vs. \textit{efficiency}. Recent works bridge the two through \textit{approximate CC} (ACC), combining redundancy, compression, and approximation. Specifically, ACC arises while relaxing the requirement of exact recovery, an assumption that is often unnecessary in modern machine learning, where approximate computations typically suffice. This enables the integration of sketching techniques to trade a small loss in accuracy for significant gains in computational and communication efficiency.

The purpose of this article is to outline the foundational techniques of CC and sketching, and to present hybrid schemes that integrate both. We also note that there is a vast literature on CC, with another major theme beyond approximate computation focusing on privacy and security \cite[Chapter 4]{LA20}. The remainder of this article can be viewed through the following lens: we first present the core primitives of CC and sketching, then review exact schemes based on redundancy, and finally present approximate methods that trade exact recovery for efficiency.

Another important distinction we should point out is that while CC typically operates over fields of finite characteristic; \textit{i.e.} $\F=\F_Q$, many sketching algorithms in numerical linear algebra operate over fields of characteristic zero; \textit{i.e.} $\F\in\{\C,\R\}$. We note, however, that sketching in a broader sense is not restricted to such fields, and includes settings over finite fields and Boolean domains, \textit{e.g.} connections to coding theory and group testing. Nonetheless, practical implementations of CC operate over $\R$ or $\C$, as is shown in \ref{RS_GC}, which can lead to numerical instability. Designing numerically stable and efficient codes over $\R$ remains challenging \cite{RT21}.


\subsection{Conceptual View}

At a high level, CC and sketching methods can be understood through the interacting axes:
\begin{enumerate}[(a)]
    \item redundancy vs. compression,
    \item exact vs. approximate recovery,
    \item structure vs. randomness.
\end{enumerate}

Early CC works emphasized \textit{redundancy} to ensure exact recovery in the presence of stragglers, leading to schemes with strong reliability guarantees but increased computational and storage overhead. In contrast, sketching emphasizes \textit{compression}, reducing problem's size, at the cost of introducing controlled error.

This naturally leads to a second axis: \textit{exact} vs. \textit{approximate} computation. While exact recovery is central in coding-theoretic formulations, many modern machine learning and optimization tasks tolerate approximation. This observation has driven the development of ACC, where redundancy and compression are combined to balance accuracy and efficiency.

Finally, methods differ in whether they rely on \textit{structure} or \textit{randomness}. Structured approaches, such as polynomial codes as expanders, offer deterministic guarantees and symmetry. In contrast, randomized approaches, such as sampling, provide flexibility and simplicity, often with probabilistic guarantees.

This article can be interpreted through this lens: exact CC schemes emphasize redundancy and structure; sketching emphasizes compression and randomness; and approximate methods bridge these viewpoints by combining elements of both to navigate trade-offs between reliability, efficiency, and accuracy.

\subsection{Significance of Coded Computing}
\label{sign_CC}

Coded computing has emerged from information theory, and has been extensively studied over the past decade. By enabling efficient, secure, and erasure-tolerant distributed computation with coding techniques, it alleviates communication bottlenecks, reduces redundancy, and enhances security.

Initially developed to accelerate algebraic tasks like \textit{matrix multiplication} (MM), polynomial evaluation, and data shuffling, CC faces challenges when applied to general machine learning tasks. To address such general purpose tasks, a method called \textit{gradient coding} (GC) has been developed~\cite{TLDK17}.

\begin{figure}
  \centering
    \includegraphics[scale=.16]{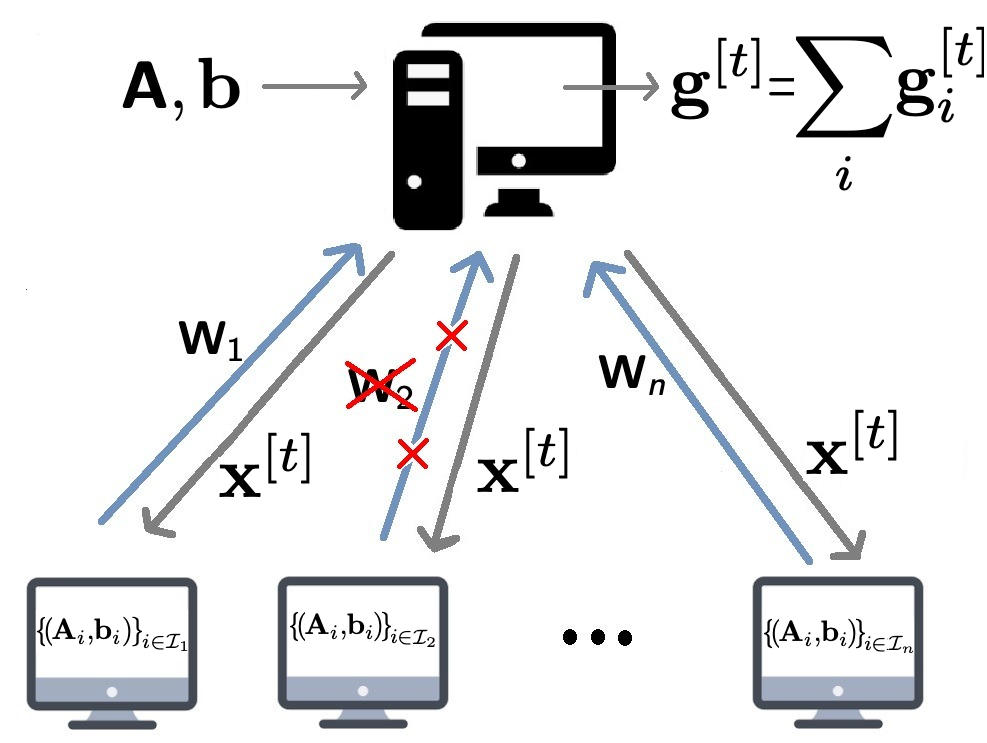}
    \caption{\underline{GC Schematic:} In an iterative first-order algorithm like gradient descent, the central server sends the current parameter vector $\xb^{[t]}$ to all $n$ servers at iteration $t$. Server $j$ holds a partition of the dataset  $\Ab$, indexed by $\I_j$, and computes the partial gradient based on its local data and $\xb^{[t]}$. The servers then encode their gradients and transmit the encoded messages to the central server. Once any $f$ encoded responses are received, the central server decodes them to recover the full gradient.
    }
  \label{gc_illustration}
\end{figure}

\begin{figure}
  \centering
    \includegraphics[scale=.16]{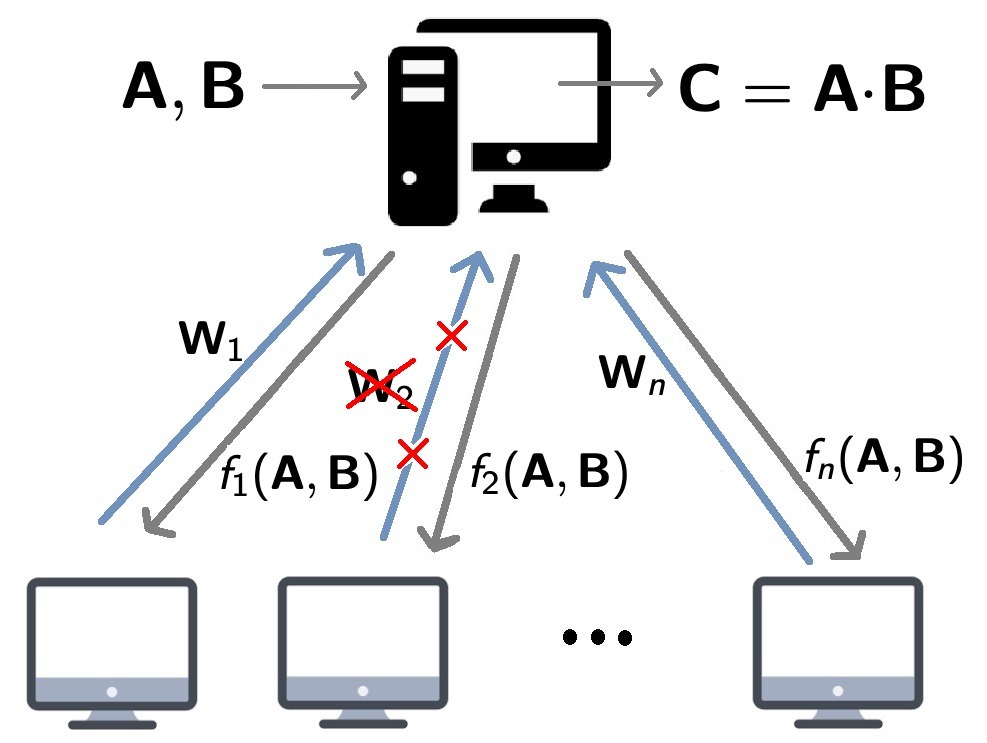}
    \caption{\underline{CMM Schematic:}
    In scenarios requiring the computation of matrix products $\Cb=\Ab\Bb$, the coordinator encodes the inputs as $\sfsty{f}_i(\Ab,\Bb)$. These are then distributed to the servers, who compute an intermediate result $\Wb_i$ based on $\sfsty{f}_i(\Ab,\Bb)$, which is sent back. Once any $f$ responses are collected, decoding permits recovery of $\Cb$.
    }
  \label{cmm_illustration}
\end{figure}

In Figures \ref{gc_illustration} and \ref{cmm_illustration} we provide schematics of GC and \textit{coded matrix-multiplication} (CMM), the two most prominent applications of CC. In a CC scheme there are generally three main steps: 1) \textit{encoding} of the data, which, depending on the application, takes place at the central server or the $n$ computational nodes, 2) \textit{computation task}, that is the distributed nodes performing a computation on the raw or encoded data, and 3) \textit{decoding}, which the central server performs as soon as $f\coloneqq n-s$ out of $n$ computations have been received, where $s$ is the number of tolerable stragglers. The parameter $f$ is known as the \textit{recovery threshold}, which is the minimum number of node responses required for a successful decoding step in CC.

Over time, as with many machine learning algorithms, it became clear that exact recovery is often an unnecessary and overly stringent requirement. Enforcing it introduces significant overhead in storage and server resources due to the redundancy inherent in erasure codes. In turn, this motivated the use of sketching in ACC.

\subsection{Significance of Sketching}
\label{sign_sketching}

As a standalone subject, sketching is also of interest to the information theory community, as it provides methods for efficiently compressing data while preserving key structures. Broadly speaking, sketching consists of dimensionality reduction techniques that balance compression with approximation accuracy, while enabling efficient storage and computation over large-scale datasets. Many such methods are rooted in the \textit{Johnson--Lindenstrauss} (JL) lemma, which states that a set of points in a high-dimensional space can be embedded into a lower-dimensional space while approximately preserving pairwise distances. This foundational result has influenced numerous areas over the past decades and is fundamental to the study of dimension reduction and geometric preservation. Since its emergence, sketching has found applications in numerical stability, matrix factorizations and decompositions, eigenvalue problems, and graph analysis.

This article focuses on the role of sketching in accelerating numerical linear algebra and optimization, especially in \textit{approximate matrix multiplication} (AMM) by leveraging matrix compression that preserves its column space with high probability. These applications tie what we will present back to CC, specifically to the subareas of CMM and GC.

In linear sketching we apply a carefully chosen oblivious or cognizant $\Sb\in\R^{ q\times N}$ on each matrix, where $q\ll N$ to reduce its size. By then applying a deterministic algorithm to the sketched surrogate problem, the overall complexity of running the algorithm is reduced, at the cost of recovering a high quality approximation. For instance, when computing the product of $\Ab\in\R^{L\times N}$ and $\Bb\in\R^{N\times M}$, we apply $\Sb$ to obtain the approximation $(\Ab\Sb)^\top(\Sb\Bb)\approx\Ab\Bb$, which is quantified in terms of the residual norm $\|\Ab\Bb-\Ab(\Sb^\top\Sb)\Bb\|_\xi$ for $\xi\in\{2,F\}$ respectively denoting the $\ell_2$ and Frobenius norms.

\begin{figure*}[t]
    \centering
    \includegraphics[width=\textwidth]{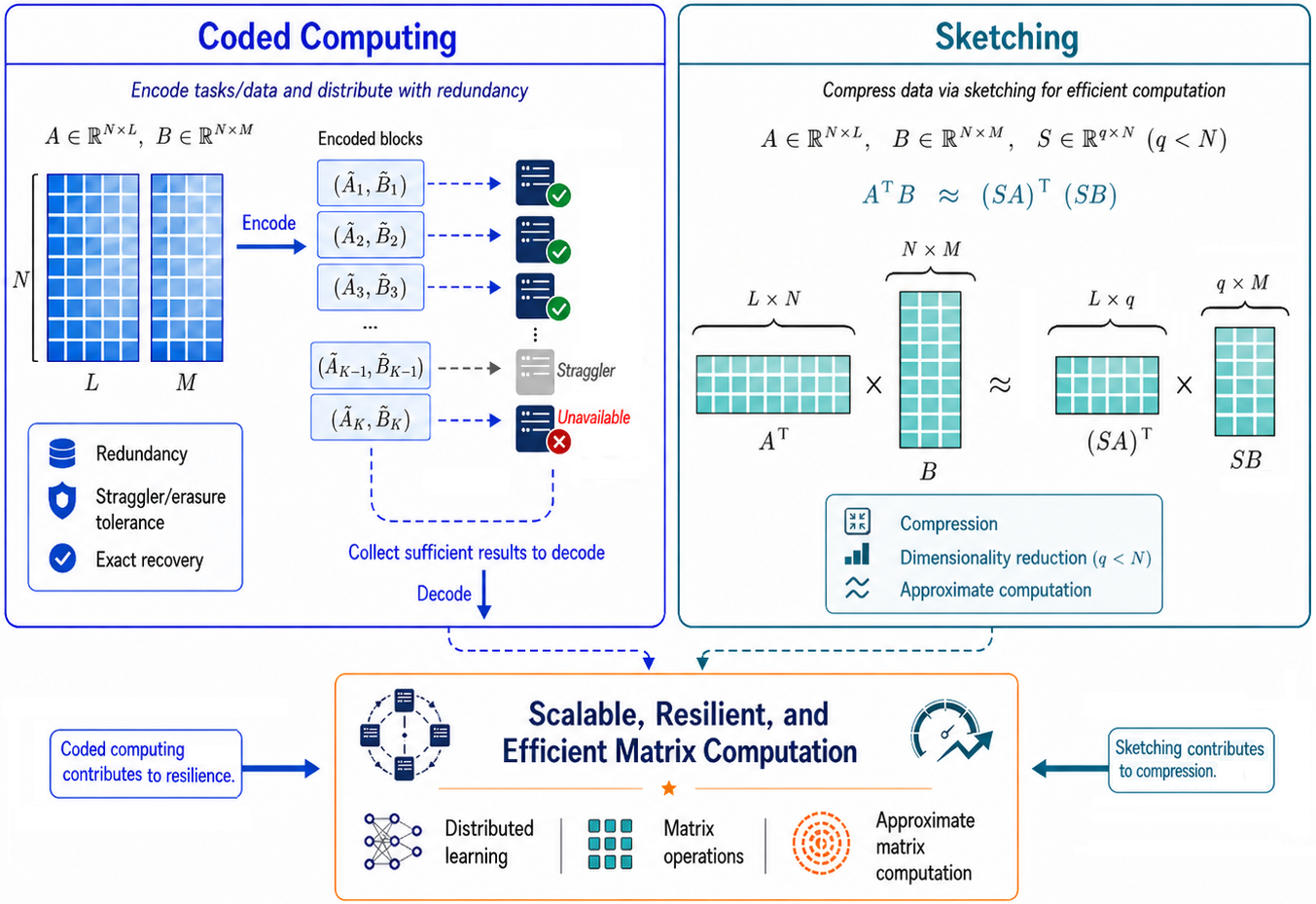}
    \caption{Overview of CC and randomized sketching for large-scale matrix computation. 
    Coded computing adds redundancy across distributed servers to tolerate stragglers and erasures, whereas sketching compresses the data to reduce computational cost, for instance through the approximation $\Ab^\top \Bb \approx (\Sb\Ab)^\top(\Sb\Bb)$, where $\Sb\in\R^{q\times N}$ and $q\ll N$. 
    The combination of these viewpoints provides a foundation for scalable, resilient, and approximate coded matrix computation.}
    \label{cc_sketching_overview}
\end{figure*}

From a conceptual standpoint, the sections that follow explore how redundancy and compression are used to address scalability in distributed systems. Section \ref{Prelim_sec} introduces the core primitives underlying coded computing and sketching. Section \ref{CC_sec} focuses on exact coded computing, where redundancy ensures reliable recovery in the presence of stragglers. In contrast, Sections \ref{AGC_sec} and \ref{ACMM_sec} consider approximate regimes, where strict recovery is relaxed and sketching enables improved efficiency. This progression reflects the central trade-off between reliability and computational cost.

To provide a unifying perspective, Table~\ref{tab_overview} summarizes the main approaches we discuss, highlighting core ideas and trade-offs between redundancy, compression, and approximation.

\begin{table*}[t]
\centering
\begin{tabular}{ |l||l|l| }
\hline
\multicolumn{3}{|c|}{\textbf{Conceptual Overview of Coded Computing and Sketching Approaches}} \\
\hline
\multicolumn{1}{|c||}{\textbf{Category}} & 
\multicolumn{1}{c|}{\textbf{Core Idea}} & 
\multicolumn{1}{c|}{\textbf{Trade-off}} \\
\hline
\hline
{\small \textbf{Exact GC \& CMM}} & {\small Redundancy via coding} & {\small Exact recovery, high overhead} \\
\hline
{\small {\textbf{Graph-based AGC}}} & {\small {Expanders, random graphs}} & {\small {Reduced redundancy, small error}} \\
\hline
{\small {\textbf{Design-based AGC}}} & {\small {Block designs, structured codes}} & {\small {Symmetry, limited flexibility}} \\
\hline
{\small {\textbf{Sketching}}} & {\small {Randomized compression}} & {\small {Fast computation, approximation error}} \\
\hline
{\small {\textbf{Sampling \& Weighting}}} & {\small Leverage scores, $CR$-CMM} & {\small Improved accuracy vs. complexity} \\
\hline
{\small \textbf{Iterative Sketching}} & {\small Fresh randomness per iteration} & {\small Reduced bias, no decoding} \\
\hline
{\small \textbf{Hybrid (Sketching + Coding)}} & {\small {CodedSketch, OverSketch}} & {\small Lower cost, approximate recovery} \\
\hline
{\small \textbf{Compression-informed CC}} & {\small Exploit the input's structure} & {\small Reduced threshold, model-dependent} \\
\hline
\end{tabular}
\caption{Summary of the main approaches discussed in this article, highlighting their core ideas and trade-offs between redundancy, compression, and approximation.}
\label{tab_overview}
\end{table*}

\section{Preliminary Background and Setup}
\label{Prelim_sec}

\subsection{Gradient Coding}
\label{GC}

The GC problem was first considered in~\cite{TLDK17}, which proposed a construction based on ``\textit{Fractional Repetition Codes}'' (FRCs). Consider a central server with a dataset $\D=\left\{(\Ab_{(i)},b_i)\right\}_{i=1}^N\subsetneq \R^d\times\R$ of $N$ samples, where $\Ab_{(i)}$ represents the features and $b_i$ the label of the $i^{th}$ sample, who can distribute $\D$ among $n$ other computational server nodes, to solve:
\begin{equation}
\label{th_star_pr}
  {\xb^{\star} = \arg\min_{\xb\in\R^d}\big\{ L(\D;\xb) \big\}}
\end{equation}
where $L(\D;\xb) = \sum_{i=1}^N \ell(\Ab_{(i)},b_i;\xb)$ is a predetermined loss function. A common approach to solving \eqref{th_star_pr} is to employ gradient descent.

The central server distributes dataset $\D$ with a certain level of redundancy, to recover the gradient based on $\D$, in the presence of stragglers. At first, $\D$ is partitioned into $k$ parts $\{\D_j\}_{j=1}^k$ each of size $N/k$. The gradient is $g\coloneqq \nabla_{\xb}L(\D;\xb) = \sum_{j=1}^k g_j$, whose summands $g_j\coloneqq\nabla_{\xb}\ell(\D_j;\xb)$ are referred to as \textit{partial gradients}.

In the distributed setting, each server returns a linear combination of its assigned partial gradients. Since stragglers are present, the central server receives a set of only $f$ completed tasks; indexed by $\I\subsetneq [n] \coloneqq\{1,\ldots,n\}$, ignoring the slowest $s$ nodes. Upon receiving any set of $f$ responses, the server decodes the encoded partial gradients to recover $g$.

Gradient coding is comprised of an encoding $\Gb\in\R^{n\times k}$, and a decoding $\ab_{\I}\in\R^f$ or $\ab_{\I}\in\R^n$ depending on the scheme; determined by $\I$. Each row of $\Gb$ represents a server's encoding vector with support size $w$, and each column corresponds to a data partition $\D_j$, each assigned to $\dt$ servers. The submatrix  $\Gb_\I\in\R^{|\I|\times k}$ includes only the rows indexed by  $\I$, and $\Gb_{ij}\neq0$ if and only if $\D_j$ is assigned to server $i$.

The $i$th server, $i\in [n]$, is assigned a subset $\J_i\subsetneq [k]$ of $w$ partial gradients to compute. The servers return an encoded linear combination of the partial gradients corresponding to their assignments. For $\gb\in\R^{k\times d}$ be the matrix whose rows are the transposed partial gradients $g_i^\top$, the central server receives encodings comprised of the rows of $\Gb_\I\gb$. The gradient $g$ of \eqref{th_star_pr} is then recoverable by applying $\ab_{\I}$:
\begin{equation}
\label{GC_criterion}
  g^\top=\ab_{\I}^\top(\Gb_{\I}\gb)=\bold{1}_{1\times k}\gb= \sum\nolimits_{j=1}^kg_j^\top.
\end{equation}
Hence, the encoding matrix $\Gb_{\I}$ must satisfy $\ab_{\I}^\top\Gb_{\I}=\bold{1}_{1\times k}$ for \textit{all} ${{n}\choose{s}}$ possible index sets $\I$. Once $g$ is obtained, an iteration of the gradient based descent algorithm is performed.

\subsection{Coded Matrix Multiplication}
\label{CMM}

Similar to GC, CMM~\cite{LSR17} introduces redundancy in distributed systems to ensure recovery in the presence of stragglers. Depending on the scheme, in CMM, each server either receives encoded data; which it computes over, or performs local computations pre-encoding. A key advantage of CMM is that encoding can be applied to $\Ab$ and $\Bb$ at the central server before distributing the tasks, as illustrated in Fig.~\ref{cmm_illustration}, offering improved recovery thresholds and greater flexibility in code design compared to GC.

In contrast, depending on $L(\D;\xb)$, GC generally does not permit pre-computation encoding. For general loss functions, encoding must occur post-computation. However, for the least squares loss, approximation schemes permit pre-computation encoding. Moreover, in a GC setup with homogeneous data assignment, and differentiable and additively separable losses, a pigeonhole argument shows that the optimal recovery threshold is $n-s$. In contrast, CMM, the recovery threshold is more difficult to characterize analytically due to uneven data partitioning and nonuniform encoding structures.

\subsection{RandNLA and Sketching}
\label{randnla_sec}

There are numerous surveys on RandNLA and sketching. An accessible monograph with clear pseudocode is~\cite{Wan15}. For more in-depth proof techniques and a comprehensive contemporary overview, see~\cite{murray2023}. Throughout this article, we focus on two key applications: AMM and \textit{$\ell_2$-subspace embedding} ($\ell_2$-s.e.).

Given $\Ab \in \F^{L \times N}$ and $\Bb \in \F^{N \times M}$, we aim to approximate $\Ab\Bb$ via a \textit{sketch} $\Ab(\Sb^\top\Sb)\Bb\approx \Ab\Bb$, where $\Sb\in\R^{q\times N}$ is an appropriate sketching matrix. For this to hold exactly, we need $\Sb^\top\Sb=\Ib_N$. This though is not possible, as $\rk(\Sb^\top\Sb)\leqslant q\ll N$. Therefore, we resort to matrices which are isotropic in expectation, \textit{i.e.} $\E[\Sb^\top\Sb]=\Ib_N$. When $\F = \R$, the ``$\bmm$'' algorithm (Algorithm~\ref{CR_alg} below) samples column-row pairs of $\Ab$ and $\Bb$ \textit{with replacement} (w.r.), using probabilities proportional to the product of their norms. We denote by $\{\pi_i\}_{i=1}^N$ and $\{p_i\}_{i=1}^N$ arbitrary and concrete sampling distributions, respectively. Let $\Mb_{(i)}$ and $\Mb^{(j)}$ denote the $i^{\text{th}}$ row and $j^{th}$ column of a matrix $\Mb$. Then, we sample $q\ll N$ row pairs $(\Ab^{(i)}, \Bb_{(i)})$ where $\Ab\Bb=\sum_{j=1}^N\Ab^{(j)}\Bb_{(j)}$, according to the distribution:
\begin{equation}
\label{cr_distr}
  p_i = \|\Ab^{(i)}\|\cdot\|\Bb_{(i)}\| \big/ \phi \qquad \quad
\end{equation}
where $\phi \coloneqq \sum_{j=1}^N \|\Ab^{(j)}\| \cdot \|\Bb_{(j)}\|$.

For $\Scal$ a multiset of $q$ sampled indices, with possible repetitions, the approximate product is: $\Ab\Bb \approx \sum_{j \in \Scal} \frac{\Ab^{(j)}}{\sqrt{q p_j}} \cdot \frac{\Bb_{(j)}}{\sqrt{q p_j}}$, and the sketches $\Abwh\coloneqq \Ab\Sb^\top$ and $\Bbwh\coloneqq \Sb\Bb$, satisfy $\|\Ab\Bb-\Abwh\Bbwh\|_F=O(\|\Ab\|_F\|\Bb\|_F/\sqrt{q})$. If we do not know the exact $\{p_i\}_{i=1}^N$ but an approximation $\{\pt_i\}_{i=1}^N$ where $\pt_i\geqslant\beta p_i$ for all $i$ and $\beta\in(0,1]$, the bound is obtained by oversampling by $1/\beta$.

The main proof technique used to prove guarantees that the above algorithm ensures $\big\|\Ab\Bb - \Abwh\Bbwh\big\|_2 \leqslant \epsilon\|\Ab\|_2\|\Bb\|_2$,  use Bernstein or Chernoff type concentration inequalities. A formal such statement is provided in Theorem \ref{complex_CR_thm}. Alternative proofs minimize the variance of the estimate $\Abwh\Bbwh$ when $\xi=F$, by showing optimality of \eqref{cr_distr}.

The second application of interest is preserving the subspace spanned by the columns of $\Ab\in\F^{N\times d}$, where $N\gg d$. This is captured by the $\ell_2$-s.e. property:
\begin{equation}
\label{subsp_emb_id}
 \|\Ib_d - \Ubwh^\herm \Ubwh\|_2 \leqslant \epsilon \quad \text{for} \quad \Ubwh \coloneqq \Sb \Ub
\end{equation}
where $\Ub\in\F^{N\times d}$ is a left orthonormal basis of $\Ab$, and $\Sb\in\R^{q\times N}$ where $N\gg q\geqslant d$, which is equivalent to preserving norms:
\begin{equation}
\label{l2_se_equiv}
  (1-\epsilon)\|\yb\|\leqslant\|\Sb\yb\|\leqslant(1+\epsilon)\|\yb\|
\end{equation}
also known as an \textit{approximate isometry} or \textit{isometric embedding}. By $\Ubwh^\herm$, we denote the  Hermitian transpose of the sketched basis $\Ubwh$.

Optimal sampling for an $\ell_2$-s.e. is through the \textit{leverage scores} $\{\ell_i\}_{i=1}^N$ of $\Ab\in\R^{N\times d}$, which are defined as $\ell_i\coloneqq\|\Ub_{(i)}\|^2$. These scores identify the most influential rows for tasks such as linear regression, and their corresponding sampling distribution is:
\begin{equation}
\label{lvg_scores_distr}
  p_i\coloneqq\ell_i\big/\sum\nolimits_{j=1}^N\ell_j=\ell_i/d.
\end{equation}
which extends to the complex case $\F=\C$.

We conclude this section by outlining a general pseudocode for sketching algorithms based on row sampling \textit{with replacement}. Choosing $\{\pi_i\}_{i=1}^N$ according to either \eqref{cr_distr} or \eqref{lvg_scores_distr}, we respectively obtain the $\bmm$ and leverage score sampling algorithms.

\begin{algorithm}[h]
\label{row_sampling_alg}
\SetAlgoLined
 \KwIn{$\Ab\in\F^{N\times d}$, distribution $\{\pi_\iota\}_{\iota=1}^N$}
 \KwOut{$\Abwh\coloneqq\Sb\Ab\in\F^{q\times d}$}
 \textbf{Initialize}: $\Sb=\bold{0}_{q\times N}$\\
 \For{$j \gets 1 $ to $q$}
   {
     {\small sample w.r. $i_j\gets\N_N$, according to $\{\pi_\iota\}$}\\
     $\Sb_{j,i_j} = 1/\sqrt{q\pi_{i_j}}$
   }
 \caption{Row Sampling}
\end{algorithm}

\subsection{Sketching for Subspace Embeddings}
\label{CounSketch_sec}

Next, we turn to subspace embeddings, which preserve a matrix’s column space and are typically built through random projections and/or row-sampling. We summarize the main constructions of such matrices $\Sb$, which reduce the ambient dimension of $\Ab\in\R^{N\times d}$ from $N$ to $q$.

\underline{\textit{Gaussian Sketch}}: Closely related to the JL-lemma, it applies a JL-transform on the left of $\Ab$. It is defined though $\bold{\Theta}$ where $\bold{\Theta}_{ij}\sim\mathcal{N}(0,1)$, which is then rescaled to obtain $\Sb=\frac{1}{\sqrt{q}}\bold{\Theta}$.

\underline{\textit{Leverage Score Sampling}}: This method is cognizant and sensitive to the geometry of $\Ab$. It is constructed via Algorithm~\ref{row_sampling_alg}, using the leverage score sampling distribution \eqref{lvg_scores_distr}.

\underline{\textit{Subsampled Hadamard Transform (SRHT)}}: Similar to the Gaussian sketch, the SRHT is also oblivious and has ties to the JL-lemma. It is comprised of three matrices, a uniform sampling and rescaling matrix of $q$ rows; $\Omb$, the normalized Hadamard matrix of order $2^{\floor*{\log_2(N)}}$; $\Hbh_N$, and a random signature matrix; $\Db$. Specifically, the SRHT is defined as: $\Sb=\Omb\cdot\Hbh_N\cdot\Db\in\{\rpm1/\sqrt{N}\}^{q\times N}$.

\underline{\textit{CountSketch}}: Among the many equivalent variants of CountSketch, the ``map-reduce'' style implementation is well-suited for distributed settings, and is outlined in Algorithm~\ref{alg_countsketch}. Given $\Ab \in \mathbb{R}^{N \times d}$, each column is hashed uniformly to one of $q$ buckets, multiplied by a uniformly random $\pm1$ sign, and summed with the other columns mapped to the same bucket.

\begin{algorithm}[h]
\label{alg_countsketch}
\SetAlgoLined
 \KwIn{$\Ab\in\R^{N\times d}$}
 \KwOut{$\Abwh\in\R^{q\times d}$}
 \textbf{Initialize}: $\Abwh=\bold{0}_{q\times d}$\\
 \For{$i \gets 1 $ to $N$}
   {
     sample $i_j$ from $\N_r$ and $s$ from $\{\rpm1\}$, each uniformly at random\\
     $\Abwh_{(i_j)}\gets \Abwh_{(i_j)}+s\cdot\Ab_{(i)}$
   }
 \caption{CountSketch}
\end{algorithm}

\subsection{A Guarantee via Matrix Concentration}

Next, we present a new representative theorem for $\bmm$ over $\F=\C$. Algorithm~\ref{CR_alg} samples $q$ pairs $(\Ab^{(i)}, \Bb_{(i)})$ from $\Ab\in\F^{ L\times N}$, $\Bb\in\F^{N\times M}$ w.r. according to \eqref{cr_distr}. The sketch is $\Ab(\Sb^\top\Sb)\Bb$, where $\Sb\in\R^{q\times N}$ has exactly one nonzero entry per row.

\begin{algorithm}[h]
\label{CR_alg}
\SetAlgoLined
 \KwIn{$\Ab\in\F^{N\times L}$ and $\Bb\in\F^{N\times M}$}
 Run Algorithm \ref{row_sampling_alg} with $\{\pi_\iota\}_{\iota=1}^N$ from \eqref{cr_distr}\\
 Obtain $\Sb\in\R_{\geqslant0}^{q\times N}$\\
 \KwOut{$(\Sb\Ab)^\herm(\Sb\Bb)\in\F^{L\times M}$}
 \caption{$\bmm$}
\end{algorithm}

\begin{Thm}
\label{complex_CR_thm}
Let $\epsilon,\delta>0$, and $\Sb\in\R_{\geqslant0}^{q\times N}$ be the sketching matrix of Algorithm \ref{CR_alg}. Then, for $q =\Theta\left(\frac{\log\left((M+L)/\delta\right)}{\epsilon^2}\cdot\Psit\right)$ where $\Psit$ is a constant depending on $\Ab$ and $\Bb$, we have:
\begin{equation}
\label{eucl_CR_prob}
  {\small\Pr\left[\frac{\|\Ab\Bb-\Ab(\Sb^\top\Sb)\Bb)\|_2}{\|\Ab\|\cdot\|\Bb\|}\leqslant\epsilon\right] > 1-\delta.}
\end{equation}
\end{Thm}



Theorem \ref{complex_CR_thm} extends to block-wise sampling~\cite{CT19,CPH20c,YM23}, where $\Ab \in \F^{L \times N}$ and $\Bb \in \F^{N \times M}$ are partitioned across the \textit{columns} of $\Ab$ and the \textit{rows} of $\Bb$, respectively. That is:
\begin{equation}
\label{part_data_1st}
  {\small \Ab=\Big[\Ab^1 \cdots \Ab^k\Big], \ \Bb=\Big[\Bb_1^\top \cdots \Bb_k^\top\Big]^\top}
\end{equation}
where $\Ab^i \in \F^{L \times \tau}$ and $\Bb_i \in \F^{\tau \times M}$, with $\tau = N/k$ for $k \in \Z_+$. The sampling distribution is defined over block-pairs $\{(\Ab^i, \Bb_i)\}_{i=1}^k$ according to $p_i \propto \|\Ab^i\|_F \cdot \|\Bb_i\|_F$. The only difference is that the sample complexity will now have a factor of $1/\tau$.

\begin{figure}
  \centering
    \includegraphics[scale=.18]{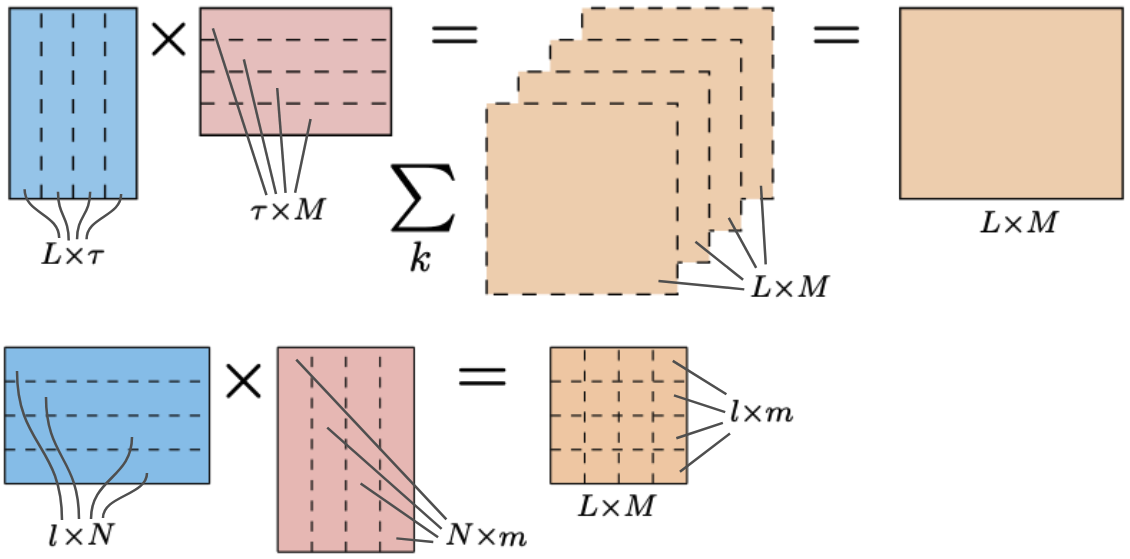}
    \caption{\underline{MM Schematics:}
    partitioning by \eqref{part_data_1st} and \eqref{part_data_2nd}.}
  \label{MM_figures}
\end{figure}

Some CC schemes adopt the partitioning:
\begin{equation}
\label{part_data_2nd}
  {\small\Ab=\Big[\Ab_1^\top \cdots \Ab_k^\top\Big]^\top,\  \Bb=\Big[\Bb^1 \cdots \Bb^k\Big]}
\end{equation}
which results in different trade-offs in recovery threshold and bandwidth. In \eqref{part_data_1st}, each product yields $\Ab^i\Bb_i\in\F^{L\times M}$, whereas in \eqref{part_data_2nd}, it yields $\Ab_j\Bb^j\in\F^{l\times m}$, where $l=L/k$, $m=M/k$ -- a visualization is provided in Fig.~\ref{MM_figures}. These trade-offs are studied in~\cite{THRD21}, which also considers AMM through unequal error protection codes.

\section{Exact CC Schemes}
\label{CC_sec}

We now turn to exact CC schemes, which aim to guarantee perfect recovery of the desired computation despite the presence of stragglers. These methods achieve robustness by introducing structured redundancy through coding, often at the cost of increased computation and storage overhead. The constructions presented in this section form the baseline upon which approximate schemes are developed.

As most ACC schemes are based on exact recovery schemes, we briefly discuss exact CC in this section. We start off with polynomial based GC methods, which fall into two categories:
\begin{enumerate}[label=(\arabic*)]
  \item \underline{\textit{Fixed Coefficient Decoding}}: uses a predetermined linear combination of the received computations,
  \item \underline{\textit{Optimal Coefficient Decoding}}: solves a least squares problem to determine $\ab_\I$.
\end{enumerate}

Similarly, CMM methods for computing $\Cb=\Ab\Bb$ are classified based on how they recover $\Cb$:
\begin{enumerate}[label=(\arabic*)]
  \item \underline{\textit{Coefficient Based}}: $\Cb$ corresponds to a coefficient of the decoding polynomial,
  \item \underline{\textit{Point Based}}: $\Cb$ is reconstructed from evaluations of the decoding polynomial at two new points.
\end{enumerate}

The key in CMM is to encode data so that the desired output is a coefficient or evaluation on a polynomial curve, enabling efficient decoding, \textit{e.g.} via the Berlekamp-Welch algorithm.

Before presenting a plethora of CMM schemes, we first detail a simpler GC scheme from~\cite{HASH17}, which illustrates key polynomial coding ideas. Here, encoding is done by the server nodes after the computations take place, and the scheme relies on Lagrange interpolation; a fundamental tool in polynomial error correction.

\subsection{Balanced Reed-Solomon Codes for GC}
\label{RS_GC}

Through Lagrange interpolation, \cite{HASH17} constructs a ``\textit{Balanced Reed-Solomon}'' ($\brs$) encoding matrix, in which each column $\Gb^{(j)}$ corresponds to a polynomial $p_j(\x)$, evaluated at a set of points $\Bcal=\{\gamma_j\}_{j=1}^n$, where $\gamma_i$ is assigned to the $i^{th}$ server. That is, $\Gb_{ij}=p_j(\gamma_i)$ for $(i,j)\in\N_n\times\N_k$, where $p_j(\x)$ are defined over a finite field $\F_Q$ and satisfy $\text{deg}(p_j)\leqslant\|\Gb^{(j)}\|_0$. To construct these polynomials, one first determines a \textit{mask matrix} $\Mb\in\{0,1\}^{n\times k}$~\cite[Alg.1]{HASH17} that is both \textit{sparsest} and \textit{balanced}, enabling the sparsest possible $\Gb$ for the fixed parameters $n,s$ and $k$. Given $\Mb$, the polynomials $\{p_j(\x)\}_{j=1}^k$ are then constructed to satisfy $p_j(\gamma_i)=0$ if and only if $\Mb_{ij}=0$, ensuring that the sparsity pattern of $\Gb$ matches that of $\Mb$. The polynomials corresponding to each column $\Gb^{(j)}$ are:
\begin{equation*}
\label{lagr_polys}
  p_j(\x) \coloneqq \prod\limits_{i:\Mb_{ij}=0}\left(\frac{\x-\gamma_i}{\gamma_j-\gamma_i}\right) = \sum\limits_{\iota=1}^{k}p_{j,\iota}\x^{\iota-1}.
\end{equation*}

The points comprising $\Bcal$ must be distinct to guarantee unique reconstruction, and are typically chosen as powers of a primitive element in $\F_Q$, to maximize distance and enable efficient correction. For our setting, we can select a primitive generator of $(\F_Q^*,\cdot)$ and set $\gamma_j=\gamma^j$ for each $j$. Since the data lies in $\R$, we may identify $\F_Q^*$ with a subgroup of the complex circle group.

This encoding matrix $\Gb$ admits the decomposition $\Gb=\Hb\Pb$, where $\Hb\in\F_Q^{n\times k}$ is Vandermonde with $\Hb_{ij}=\gamma_i(j-1)$ and $\Pb_{ij}=p_{j,i}$ for $\Pb\in(\F_Q^*)^{k\times k}$. Furthermore, $p_j(\x)$ is defined by $\Pb^{(j)}$, and evaluates as $p_j(\gamma_i) = \sum_{\iota=1}^{k}p_{j,\iota}\gamma_i^{\iota-1} = \langle\Hb_{(i)},\Pb^{(j)}\rangle$. A normalization is applied so that the constant term of every polynomial is $1$, \textit{i.e.} $\Pb_{(1)}=\vec{\bold{1}}$. This ensures that for any $\I$, the decoding vector $\ab_\I$ is simply $\big(\Hb_\I^{-1}\big)_{(1)}$, which is constructable in $\ow(f^2)$.
To visualize the resulting encoding matrix, we provide a concrete example with parameters $n=8$, $k=4$, $\tilde{d}=6$, and $w=3$:
$$ \Gb = \begin{pmatrix}
p_1({\gamma_1}) & p_2({\gamma_1}) & p_3({\gamma_1}) & {\gray 0} \\

p_1({\gamma_2}) & p_2({\gamma_2}) & p_3({\gamma_2}) & {\gray 0} \\

p_1({\gamma_3}) & p_2({\gamma_3}) & {\gray 0} & p_4({\gamma_3}) \\

p_1({\gamma_4}) & p_2({\gamma_4}) & {\gray 0} & p_4({\gamma_4}) \\

p_1({\gamma_5}) & {\gray 0} & p_3({\gamma_5}) & p_4({\gamma_5}) \\

p_1({\gamma_6}) & {\gray 0} & p_3({\gamma_6}) & p_4({\gamma_6}) \\

{\gray 0} & p_2({\gamma_7}) & p_3({\gamma_7}) & p_4({\gamma_7}) \\

{\gray 0} & p_2({\gamma_8}) & p_3({\gamma_8}) & p_4({\gamma_8}) \\

 \end{pmatrix} . $$

\subsection{MatDot Codes for CMM}
\label{MatDot_sec}

``MatDot CMM''~\cite{DFHJCG19} exploits \eqref{part_data_1st}, to define encoding polynomials $p_{\Ab}(\x)=\sum_{j=1}^{k}\Ab^j\x^{k-1}$ and $p_{\Bb}(\x)=\sum_{j=1}^{k}\Bb_j\x^{k-j}$, which are evaluated at distinct elements $\{\gamma_i\}_{i=1}^n\subsetneq\F_Q$; as in \cite{HASH17}. Each server $i$ receives $p_{\Ab}(\gamma_i)$ and $p_{\Bb}(\gamma_i)$ and returns the degree $2(k-2)$ product polynomial $C(\gamma_i)=p_{\Ab}(\gamma_i)p_{\Bb}(\gamma_i)$. From this, the coefficient of $\x^{k-1}$ in $p_{\Ab}(\x)p_{\Bb}(\x)$ is the desired product $\Ab\Bb=\sum_{j=1}^k\Ab^j\Bb_j$.

Once any $2k-1$ evaluations of $C(\x)$ are received, polynomial interpolation or Reed-Solomon decoding recovers $\Ab\Bb$. Follow up work also showed that this recovery threshold of $2k-1$ is optimal. As we will see, this construction has also been extended to \textit{approximate CMM} (ACMM) via sketching~\cite{CT19,CPH20c}.

\subsection{Polynomial Codes for CMM}
\label{poly_codes}

Unlike MatDot Codes, ``Polynomial CMM Codes''~\cite{YMAA17} rely on the partitioning \eqref{part_data_2nd}, where $\Ab$ and $\Bb$ are divided into row and column blocks, respectively. The servers compute the product of their encoded blocks and return the result, and the coordinator's decoding is done via a Vandermonde matrix inversion once enough responses are collected.

Given $a,b\in\Z_+$, the $(a,b)$-\textit{polynomial code} is defined by $\Abt^a_i=\sum_{j=1}^k\Ab_j\gamma_i^{(j-1)a}$ and $\Bbt^b_i=\sum_{j=1}^k\Bb^j\gamma_i^{(j-1)b}$, and each server $i$ computes $\Wb_i = \Abt^a_i\cdot\Bbt^b_i = \sum_{j=1}^k\sum_{l=1}^k \Ab_j\Bb^j\gamma_i^{(j-1)a+(l-1)b}$. To ensure recovery from any $k^2$ responses, $(a,b)$ are chosen so that all exponents are distinct. A canonical choice is $(a,b)=(1,k)$, yielding $h(\x) = \sum_{j=1}^k\sum_{l=1}^k\Ab_j\Bb^l\x^{(j-1)+(l-1)k}$, where the coefficient of each monomial corresponds to a unique submatrix of $\Cb=\Ab\Bb$. With distinct evaluation points $\{\gamma_i\}_{i=1}^n$, any $k^2$ evaluations suffice to interpolate $h(\x)$ and recover $\Cb$, since  $\text{deg}(h)=k^2-1$.

As an example, let $n=5$ and $s=1$. Similar to $\brs$ GC, both encoding and decoding here rely on the Vandermonde structure of the evaluation points. Each server computes $\Wb_i=\Ab_1\Bb^1+\gamma_i\Ab_2\Bb^1+\gamma_i^2\Ab_1\Bb^2+\gamma_i^3\Ab_2\Bb^2$, by evaluating a degree 3 matrix polynomial at $\gamma_i$. A system defined by a Vandermonde matrix $\bold{\Gamma}=\big(\gamma_i^{j-1}\big)$ for $(i,j)\in\N_n\times\N_{(n-s)}$ is formed by appending the $\Wb_i$'s. A single straggler corresponds to erasing its corresponding row in $\bold{\Gamma}$, but the remaining system is square and full-rank due to the MDS property of $\bold{\Gamma}$. Hence, $\Cb$ is recoverable by directly inverting this reduced system, ensuring decodability.

\subsection{Entangled Polynomial Codes for CMM}

``Entangled Polynomial Codes''~\cite{YMAA20} generalize and improve upon PolyDot Codes~\cite{DFHJCG19} by enabling arbitrary partitioning of the input matrices while achieving a strictly lower recovery threshold, by a factor of 2. This improvement is realized through the use of structured redundancy that is specifically designed to exploit the bilinear nature of matrix multiplication. In the case where $\Ab$ and $\Bb$ are respectively partitioned across their rows and columns by $k_1$, $k_2$ and $k_2$, $k_3$, then Entangled Polynomial Codes achieve a recovery threshold of $k_1k_2(k_3+1)-1$.

Similar to a $(a,b)$-polynomial codes~\cite{DFHJCG19}, Entangled Polynomial Codes introduce an additional parameter $\theta$, giving rise to the notion of a $(a,b,\theta)$-\textit{polynomial code}. Rather than presenting the full construction explicitly, we illustrate the concept with a simplified example, to highlight the role of entanglement.

Consider a scenario where $n=5$ and $s=1$. In the corresponding encodings of the Entangled Polynomial Code, both $\Ab$ and $\Bb$ are partitioned into two across one dimension (for this simple example, we do not partition across the other dimension). Each server $j\in\N_5$ receives their two corresponding coded submatrices $\Abt_i\coloneqq\Ab_0+\gamma_i\Ab_1$ and $\Bbt_i\coloneqq \gamma_i\Bb^0+\Bb^1$, and computes:
$$ \Abt_i\Bbt_i = \Ab_0\Bb^1+\gamma_i(\Ab\Bb^0+\Ab_1\Bb^1)+\gamma_i^2\Ab_1\Bb^0 $$
which when aggregated comprise the system whose Vandermonde matrix is of size $5\times 3$. Therefore, any two erasures are permissible and we will have an invertible square system, \textit{i.e.} we can tolerate $s=2$ stragglers.

\section{Approximate Gradient Coding}
\label{AGC_sec}

While the schemes in the previous section ensure exact recovery, such guarantees are often stronger than necessary in many large-scale learning and optimization tasks. Enforcing exactness can introduce significant redundancy and computational overhead, motivating the consideration of approximate alternatives. In this section we turn to such schemes, where recovery constraints are relaxed and ideas from sketching are incorporated to achieve improved efficiency while maintaining controlled error.

This area, of ACC, is vast and rapidly evolving, and has gained significant traction following advances in exact CC. In this section and Sec.~\ref{ACMM_sec}, we present \textit{approximation} schemes that build upon the \textit{exact} constructions discussed thus far, highlighting their connections to sketching. We also discuss representative approaches that rely on similar principles, illustrating the diversity of techniques used in ACC. Similar to Sec.~\ref{CC_sec}, this section focuses on \textit{approximate GC} (AGC), while the next section focuses on ACMM.

In contrast to exact gradient coding, AGC relaxes the requirement of perfect recovery in favor of improved efficiency and flexibility. The goal is to tolerate stragglers while allowing for controlled approximation error, aligning naturally with principles from sketching. The schemes presented in this section illustrate how this relaxation enables new design approaches that reduce overhead while maintaining strong performance guarantees.

\subsection{Gradient Coding from Expanders}
\label{GC_expanders}

Recall that the GC condition requires co-designing an encoding-decoding pair $(\Gb, \ab_\I)$ such that $\ab_\I^\top \Gb_\I = \bold{1}_{1\times k}$ for all subsets of $\I$ of size $f$. In AGC, this condition is relaxed to $\ab_\I^\top \Gb_\I \approx \bold{1}_{1\times k}$, enabling a reduced decoding overhead at each iteration, aligning with common practices which permit approximate gradients.

The first work to consider AGC~\cite{RTTD17} formalized this relaxation, using the metric:
\begin{equation}
\label{AGC_err}
  \max_{\I\in\binom{[n]}{n - s}}\left\{ \|\ab_\I^\top \Gb_\I - \bold{1}_{1\times k}\|_2 \right\} \leqslant \epsilon
\end{equation}
for a small $\epsilon > 0$. This work, proposed setting $\Gb$ as the normalized adjacency matrix of a sparse, connected, $\Delta$-regular expander graph on $n$ nodes, with $k = n$ data partitions. The decoding vector $\ab_\I$ is then defined entry-wise as $(\ab_\I)_l = \frac{n}{n - s}$ if $l \in \I$, and zero otherwise. Under this construction, the estimate's error is bounded by $\epsilon \leqslant \frac{\lambda}{\Delta} \sqrt{\frac{ns}{n - s}}$, where $\lambda$ is a bound on the second-largest eigenvalue of $\Gb$.

To minimize $\lambda$ and improve the accuracy, Ramanujan graphs were leveraged; as they are optimal expanders that satisfy $\lambda \leqslant 2\sqrt{\Delta - 1} + o_n(1)$. These graphs can be explicitly constructed and offer near-optimal spectral gap properties, leading to tighter guarantees in \eqref{AGC_err}.

It is worth noting that additional constructions based on expanders were proposed in~\cite{GW20}, extending existing  analysis and techniques to alleviate both random and adversarial stragglers.

These constructions highlight how spectral properties of graphs can be leveraged to control approximation error, providing a structured alternative to random sketching.

\subsection{GC from Sparse Random Graphs}

Following~\cite{RTTD17}, leveraging random constructions appeared to be a promising approach for addressing the AGC problem. This was precisely what~\cite{CPE17}, as they used random graphs to design simpler codes for the random and adversarial straggler models. For a given encoding $\Gb$, they considered two decoding strategies:
\begin{enumerate}
  \item[$(i)$] {\small \underline{\textit{One-Step}}: $\ab_\I^\top=\rho\bold{1}_{1\times k}$ for $\rho$ fixed and all $\I$,}
  \item[$(ii)$] {\small \underline{\textit{Optimal}}: $\ab_\I^\star = \arg\min\big\{\|\ab_\I^\top\Gb_\I-\bold{1}_{1\times k}\|_2\big\}$.}
\end{enumerate}
where $(ii)$ yields the best possible decoding vector per iteration via $\ab_\I^\star = \bold{1}_{1\times k} \Gb_\I^\dagger$, while $(i)$ is a fixed rule.

Although binary FRCs~\cite{TLDK17} achieve low decoding error under random stragglers when using $(ii)$, their performance degrades under adversarial straggler selection. To address this,~\cite{CPE17} proposed ``Bernoulli GCs'', whose encoding is $\Gb \in \{0,1\}^{n \times k}$ with entries $\Gb_{ij} \overset{\text{iid}}{\sim} \text{Bern}(s/k)$. Compared to the Ramanujan based encodings of~\cite{RTTD17}, which offer strong guarantees but are difficult to construct, Bernoulli GCs provide a simple and practical alternative.

Moreover,~\cite{CPE17}  proved NP-hardness of the adversarial straggler selection problem for general codes, via reductions to the $k$-densest subgraph problem, motivating follow up works~\cite{KKR19,SW22}.

Constructions based on random graphs further demonstrate how randomness can simplify code design while retaining guarantees, closely mirroring the role of randomness in sketching methods.

\subsection{GC Based on Block Designs}

Another class of AGC schemes for mitigating adversarial stragglers through graph constructions, are based on \textit{balanced incomplete block designs} (BIBDs)~\cite{KKR19}, which are highly symmetric incidence structures. A $(v,b,f,r,\lambda)$-BIBD consists of $v$ points and $b$ blocks of size $k$, such that each point appears in $r$ blocks, and every pair of points is contained in exactly $\lambda$ blocks. The incidence matrix $\Gb\in\{0,1\}^{v\times b}$ where $\Gb_{ij}=1$ if and only if point $i$ is in block $j$, defines the encoding matrix $(n,k,w,\dt)$-BIBD GC by setting $n=v$ $k=b$, $w=f$, $\dt=r$, and has an optimal decoding vector of $\ab_\I^\top = \rho \bold{1}_{1\times(n-s)}$, with $\rho = \frac{w}{w + \lambda(n - s - 1)}$ for any set $\I$.

Since BIBDs exist only for limited parameter sets, subsequent work~\cite{SW22} introduced ``Soft-BIBD GC'', which relaxes the strict BIBD constraints while ensuring that $w$ and $\dt$ are satisfied in expectation. In this construction, a probability distribution $p(x^n)$ is defined over $\Gb$'s columns, which $\Gb$ is generated by sampling each of its $k$ columns i.i.d. from $p(x^n)$. Ensuring BIBD-like behavior reduces to solving a linear system $\bold{\Gamma}\pb = \bb$ for $\pb \geqslant 0$, where $\bold{\Gamma}$ is shown to be the generator matrix of an order 2 Reed–Muller code. As in~\cite{CPE17,KKR19}, this work also suggested decoding via $\ab_\I^*$. This work also proposed ``Product GC'', in which Kronecker products of simpler GC encoding matrices generate new codes. These offer scalability and reduced density while maintaining error performance similar to their component codes.

More recently,~\cite{JZLW24} introduced ``Sparse Gaussian GC'', another approximate version of BIBD GC. The encoding is $\Gb = \Xb \odot \Bb$, where $\Bb_{ij} \sim \text{Bern}(\gamma)$ and each column of $\Xb$ is sampled i.i.d. from $\mathcal{N}(\mub, \Sigb)$, with $\mub = a\bold{1}_n$ and $\Sigb = c\bold{1}_n\bold{1}_n^\top + (b - c)\Ib_n$. This construction also mimics BIBD properties in expectation, while allowing significantly more flexibility in the parameter selection.

These constructions emphasize the role of combinatorial structure in achieving balanced redundancy, offering a deterministic counterpart to randomized approaches.

\subsection{Weighted Gradient Coding}
\label{weighted_GC}

In contrast to the AGC schemes discussed thus far, the scheme in~\cite{CPH20} does not rely on random graph constructions, but instead utilizes weighting on top of sampling; with $\brs$ GC. Their goal is to integrate $\ell_2$-s.e. via \textit{block} leverage score sampling into GC, yielding approximate gradients through two stages of compression. A generalization of leverage scores \eqref{lvg_scores_distr} was defined, where if $\Ub$ is partitioned across its rows into $k$ blocks, then $\Pi_i = \|\Ub_i\|_F^2/d$ defines a distribution over the corresponding \textit{blocks} of $\Ab$. Then, sampling is performed w.r. until $r=q/\tau$ distinct blocks are selected, each rescaled by $\sqrt{1/(r\Pi_{i_j})}$.

The central server tracks which $r$ blocks were sampled and how often, which information is encoded in a weight vector $\wbt \in \Z_+^{1\times r}$. The compressed matrix $\Abt = [\Abt_1 \dots \Abt_r] \in \R^{q \times d}$ and label vector $\bbt \in \R^q$ are constructed accordingly. Unlike exact GC, the objective of this scheme is to find a pair $(\Gbt, \ab_\I)$ satisfying $\ab_\I^\top \Gbt_\I = \wbt$ for \textit{all} $\I$. To achieve this, ``Weighed AGC'' exploits the structure of the $\brs$ GC, by defining $\Gbt \coloneqq \Gb \cdot \mathrm{diag}(\wbt)$ with $k \gets r$. The decoding step is identical to that of $\brs$ GC.

Two features distinguishing this scheme are that more sampling trials take place, and the resulting gradient matches that of a sketched version of the problem, leading to superior approximation and convergence guarantees compared to naive sampling.

In the schemes of this subsection, we observe a connection making explicit the link between GC and sketching, where importance sampling plays a central role in improving approximation quality.

\subsection{AGC Based on Iterative Sketching}

As we have seen, most GC and AGC schemes rely on a decoding step and a strict recovery threshold to obtain the gradient at each iteration. Furthermore, decoding is generally expensive and becomes a bottleneck when applied repeatedly. To this end, the works of~\cite{CPH24,CMPH23} introduce schemes for linear regression via $\ell_2$-s.e. to bypass decoding altogether, while also allowing flexibility in the number of stragglers.

In~\cite{CPH24}, block leverage score sampling (without weighting) from~\cite{CPH20} was adopted, and stragglers were interpreted as erasures in a communication channel model, \textit{i.e.} straggling outputs are treated as ``erasures'', and were modeled probabilistically as in~\cite{LLPPR17}. Assuming homogeneous servers, this is equivalent to the central server uniformly sampling the requested computations at each iteration. In the encoding stage the blocks are replicated based on their block leverage scores and desired response time which are appropriately rescaled, and the central server receives a fresh sketched gradient at each iteration. Since encoding mimics updated sketching pre-computation, no decoding is necessary in order to recover a satisfactory approximation, so the server simply aggregates all available computations once enough have been received.

To incorporate security and further simplify this scheme,~\cite{CMPH23} introduced a pre-processing step where a random projection $\Pib \in \R^{N \times N}$ is applied to $\Ab$ and $\bb$, yielding $\Abt = \Pib \Ab$ and $\bbt = \Pib \bb$, which are partitioned and distributed to the $n$ servers. The application of $\Pib$ masks the raw data, ensuring privacy, while simultaneously flattening the block leverage scores, similar to the SRHT; permitting uniform sampling for an $\ell_2$-s.e. In this scheme, each block is assigned to a single server. A drawback is the potential overhead in generating and applying $\Pib$, and a trade-off emerges where faster random projections offer lower security, and vice versa.

A major benefit of these AGC approaches is that they implicitly generate a new sketch of the data at each iteration via a fresh random subset of servers, avoiding the bias of the ``sketch-and-solve'' paradigm, where a single sketch is used to approximate the surrogate problem $\xb^\star \approx \arg\min_{\xb \in \R^d} \|\Sb(\Ab\xb - \bb)\|_2^2$. This ``iterative sketching'' approach has been shown to be superior experimentally and theoretically.

Moreover, these techniques extend to Newton’s method, enabling unbiased estimation of both gradients and Hessians at each iteration. In $\ell_2$-s.e. based AGC schemes, it has also been shown that the gradient error \eqref{AGC_err} can be directly quantified by the $\ell_2$-s.e. error $\epsilon$ \eqref{subsp_emb_id}.

These approaches highlight a shift toward sketching-driven designs, where repeated randomization replaces explicit decoding while maintaining strong guarantees.

\subsection{Encoding through ETFs, and the RIP}
\label{ETFs_GC}

Another line of work, initiated by~\cite{KSD17}, introduces redundancy via \textit{equiangular tight frames} (ETFs) and partitions the system into smaller linear problems whose solutions gradients are locally computed by the computational nodes. In contrast to GC schemes~\cite{TLDK17,HASH17,RTTD17}, redundancy here is embedded directly in the optimization formulation. Inspired by sketching, the data is linearly encoded and distributed across the $n$ nodes, which solve local subproblems independently, ignoring stragglers.

The objective function of \eqref{th_star_pr} is now encoded through an ETF matrix $\Sbwt = \big[\Sbt_1^\herm \ \cdots \ \Sbt_n^\herm\big]^\herm$, where each $\Sbt_i\in \F^{\qt \times N}$ is the encoding of the $i^{th}$ node with $\qt \ll N$. The global objective becomes $L_{\Sbwt}(\Ab,\bb;\xb) \coloneqq \|\Sbwt(\Ab\xb - \bb)\|_2^2$, which is equal to the summation of the locally minimized objective functions: $\sum_{i=1}^n \|\Sbt_i(\Ab\xb - \bb)\|_2^2$. For $\I$ the set of non-stragglers, the aggregated objective is $L_{\Sbwt_{\I}}(\Ab,\bb;\xb) \coloneqq \sum_{i \in \I} \|\Sbt_i(\Ab\xb - \bb)\|_2^2$. At each iteration, the central server collects $f$ local gradients to perform a descent step.

ETFs are well-suited for this setting as they introduce redundancy, ensure maximal information diversity across vectors, and enable exact reconstruction in the absence of stragglers. The construction of $\Sbwt$ in~\cite{KSD17} uses Paley graphs, which are self-complementary strongly regular graphs defined over $\F_Q$ with $Q \equiv 1 \pmod{4}$, whose Seidel adjacency matrices yield real ETFs. Specifically, entries of the adjacency matrix $\Gb \in \{0,1\}^{Q \times Q}$ are nonzero only if the difference $i-j$ is a quadratic residue in $\F_Q$.

Similar to other constructions we have seen, this construction is limited by number-theoretic constraints, and its existence hinges on the availability of finite fields of certain orders. Despite these constraints and not knowing when Paley graphs exist, several open problems arise from generalizing their structure; or using them in areas such as coding, design and frame theory.

ETFs are also linked to the \textit{restricted isometry property} (RIP), a key concept in compressed sensing. A matrix $\Phib\in\R^{p\times q}$ where $p<q$, satisfies the order-$\kappa$ RIP if, for all $\kappa$-sparse $\xb$:
\begin{equation*}
  (1 - \delta_\kappa)\|\xb\| \leqslant \|\Phib\xb\| \leqslant (1 + \delta_\kappa)\|\xb\| 
\end{equation*}
for $\delta_\kappa \in (0,1)$, which mirrors the norm-preservation property of the $\ell_2$-s.e. condition \eqref{l2_se_equiv}. The RIP ensures that sparse vectors remain distinguishable under projection; acting like an isometry, while also resembling the $\ell_2$-s.e. property. Moreover, ETFs minimize coherence, \textit{i.e.} the maximum absolute inner product between frame vectors, making them nearly orthogonal on sparse supports. Both random and deterministic constructions satisfying RIP have been extensively studied in the literature. In this context, ETFs can serve as structured surrogates for small $\kappa$, offering deterministic constructions with behavior similar to the RIP.

These constructions illustrate how geometric properties of embeddings can be used to encode redundancy, bridging ideas from compressed sensing and coded computation.\\

\noindent\underline{\textbf{Summary of Section~\ref{AGC_sec}}}: Across the AGC schemes we presented, a unifying theme is the relaxation of exact recovery in favor of bounded approximations, enabling reduced redundancy and improved efficiency. This perspective is particularly well-aligned with modern machine learning applications, where exact computations are rarely required and often intractable at scale, especially in deep learning settings. The methods discussed reflect different trade-offs between decoding complexity, approximation error, and robustness to stragglers.

\section{Approximate Coded MM}
\label{ACMM_sec}

In contrast to AGC, which focuses on iterative optimization, ACMM targets the efficient computation of core linear algebra subroutines involving one or multiple matrix products. The goal is to reduce computational and communication costs while maintaining accurate approximations, making sketching techniques particularly effective in this setting. In this section, we review a variety of such ACMM schemes, illustrating how sampling, compression, and coding can be combined to enable scalable AMM computations.

\subsection{CMM by Random Sampling}
\label{CMM_rand_smapling}

The work of~\cite{CT19} was the first to propose ACMM, doing so by developing two schemes which combined Algorithm \ref{CR_alg} with MatDot Codes, and by through their developments determined a trade-off between approximation error and recovery threshold. Both their schemes randomly sample $r$ pairs $(\Ab^j, \Bb_j)$ from the partitioning in \eqref{part_data_1st}; indexed by $\I$, and consider the approximation error of $\Cbwh \approx \Ab\Bb$, which is quantified via bounds on $\E\big[\|\Ab\Bb - \Cbwh\|_F^2\big]$.

Their first scheme, ``Coded Set-wise Sampling'', samples $\I$ of size $r$ (without replacement) w.p. $P_\I$. The sampled submatrices are encoded through $\Abt(\x)=\sum_{j=1}^r\Ab^{\I_j}\x^{j-1}/\sqrt{cP_\I}$ and $\Bbt(\x)=\sum_{j=1}^r\Bb_{\I_j}\x^{K-j}/\sqrt{cP_\I}$, where $c = {{k}\choose{r}}\cdot r/k$. These polynomials are evaluated at distinct points $\{\gamma_i\}_{i=1}^n\subsetneq\F_Q$ which evaluations are sent to the servers, who compute $\Abt(\gamma_i)\Bbt(\gamma_i)$. By the MatDot decoding step, the approximation $\Cbwh_\I = \sum_{l=1}^r \frac{\Ab^{\I_l} \Bb_{\I_l}}{c P_\I}$ is recovered once $2r-1$ responses are collected. A caveat though is that calculating the optimal probabilities $P_\I$ costs more than computing $\Ab\Bb$, so the authors resort to uniform sampling.

Their second scheme, ``Coded Independent Sampling'', samples w.r. each index $\I_\iota \in \N_k$ independently for each $\iota \in \N_r$. The encodings are through $\Abt(\x)=\sum_{j\in\I}\Ab^j\x^{j-1}/\sqrt{rP_{\I_j}}$ and $\Bbt(\x)=\sum_{j\in\I}\Bb_j\x^{r-i}/\sqrt{rP_{\I_j}}$,
and as before; each server computes $\Abt(\gamma_i)\Bbt(\gamma_i)$. Upon receiving $2r-1$ responses, the approximation $\Cbwh_\I = \sum_{l=1}^r \frac{\Ab^{\I_l} \Bb_{\I_l}}{c P_\I}$ is recovered through MatDot decoding. In contrast to the aforementioned scheme, this approach allows repeated indices in $\I$, and the sampling complexity is reduced.

These schemes directly reflect classical sketching ideas, where sampling reduces computational cost at the expense of controlled approximation error.

\subsection{Weighted $CR$-CMM}
\label{weighted_CR}

A drawback of the schemes in~\cite{CT19} is that they either rely on expensive to compute distributions, or do not specify the distribution. To address these shortcomings, the work of~\cite{CPH20c}, which also utilizes Algorithm \ref{CR_alg} (also referred to as ``$CR$-MM'') and MatDot Codes, proposes a more practical and effective approach. Moreover, the block sampling sketching scheme developed in~\cite{CPH20c}, offers the best accuracy vs. runtime trade-off~\cite{YM23}.

The ``$CR$-CMM'' approach samples pairs $(\Ab^i,\Bb_i)$ w.p. proportional to $\|\Ab^i\|_F^2\cdot\|\Bb_i\|_F^2$, generalizing its row sampling counterpart \eqref{cr_distr}. Although the optimal distribution which is proportional to $\|\Ab^i\Bb_i\|_F^2$ yields better approximations, it is more expensive to compute than directly evaluating $\Ab\Bb$, a shortcoming of~\cite{CT19}.

The sampling in $CR$-CMM is performed w.r. until $r$ distinct block pairs have been obtained, as in Weighted GC~\cite{CPH20}. Let $\wb\in\Z_{\geqslant 0}^{1\times K}$ denote the weight vector where $\wb_i$ counts how many times pair $(\Ab^i,\Bb_i)$ was sampled. Unused block pairs are discarded, and the sampled ones are assembled into $\Abt = [\Ab^{\I_1} \ \cdots\ \Ab^{\I_r}]$ and $\Bbt = [\Bb_{\I_1} \ \cdots\ \Bb_{\I_r}]$. After rearranging the nonzero weights $\wb_{\I_j}$ to match the order in $\Abt$ and $\Bbt$, we define $\wbt \in \Z_+^{1\times r}$. MatDot coding is then applied to $(\Abt,\Bbt)$, through the modified encoding polynomials $p_{\Abt}(\x)=\sum_{j=1}^r\sqrt{\wbt_j}\cdot\Abt^j\x^{r-1}=\sum_{j=1}^r\sqrt{\wb_{\I_j}}\cdot\Ab^{\I_j}\x^{r-1}$ and $p_{\Bbt}(\x)=\sum_{j=1}^r\sqrt{\wbt_j}\cdot\Bbt_j\x^{r-j}=\sum_{j=1}^r\sqrt{\wb_{\I_j}}\cdot\Bb_{\I_j}\x^{r-j}$. The MatDot decoding step remains intact.

A key advantage of this scheme is that, for the same level of sketching compression, it performs more sampling trials than other ACMM schemes based on Algorithm \ref{cr_distr}, yielding stronger approximation guarantees. This benefit stems from the use of weighting, similar to the AGC approach of~\cite{CPH20}. Moreover, the error bound of this scheme is sharpened by the block variant of Theorem \ref{eucl_CR_prob}, improving upon the guarantees provided in~\cite{CPH20c}.

This scheme further reinforces the connection to sketching, as weighted schemes improve accuracy by providing additional compression compared to the schemes of~\cite{CT19}.

\subsection{CodedSketch}

The ``CodedSketch scheme''~\cite{JM21} presents an ACMM framework that merges Entangled Polynomial Codes and CountSketch. The core idea is to replace the exact product with a compressed approximation, using hash based sketching.

Instead of directly performing MM, CodedSketch leverages hash functions to compress rows and columns of the input matrices, representing the result through what they define as \textit{sketch polynomials}. These polynomials encode the data in a compact form, which are evaluated and multiplied at the computational servers. Using techniques from polynomial coding and Lagrange CC, the central server aggregates the responses and interpolates the final result. A median based decoding procedure is then applied to recover approximate submatrices of the product, which are aggregated to produce an approximation.

This scheme exemplifies a tight integration of coding and sketching, combining structured redundancy with hash-based compression.

\subsection{OverSketch}

The ``OverSketch'' scheme~\cite{GWCR18} introduces redundancy in distributed MM using additional CountSketches to mitigate the effect of stragglers. Specifically, it computes $\breve{\Ab}=\Ab\Sb$ and $\breve{\Bb}=\Sb^\top\Bb$ via a CountSketch matrix $\Sb$, which sketches are partitioned into $b\times b$ blocks. Each $b\times b$ block of the product $\Cb=\Ab\Bb$ is then approximated by multiplying a row-block of $\breve{\Ab}$ with a column-block of $\breve{\Bb}$, where each multiplication is assigned to a single server.

Unlike CodedSketch, OverSketch computes each block submatrix of $\Cb$ independently, and does not exploit coding gains across different blocks, \textit{i.e.} each server’s computation contributes only to a single block. Nonetheless, OverSketch has been extended to distributed Newton sketching for solving convex optimization problems, broadening its applicability.

In contrast to CodedSketch, this approach prioritizes scalability by decoupling computation across blocks, trading coding gains for simpler parallelization.

\subsection{Compression-Informed Coded Computing}
\label{compression_informed}

A distinct direction proposed in~\cite{RCHV23} utilizes linear dependencies in the input matrices for CC tasks. In prior work, the authors introduced a scheme for computing any homogeneous polynomial $g(\x)$ by constructing a univariate polynomial $p(\z)$ through scaled versions of $k$ input points. Interpolating $g(p(\z))$ allows exact recovery of $g(\cdot)$ at the original points, reducing the number of required servers compared to input-oblivious approaches. This idea was extended to exploit coordinate-wise dependencies, enabling computation even when the input points are independent in some coordinates.

In the case of CMM, the scheme of~\cite{RCHV23} applies this principle using a \textit{grid partitioning} of $\Ab$ and $\Bb$, \textit{i.e.} partition them across both rows and columns. The approximate product $\Cbwh \approx \Ab\Bb$ is constructed from sampled submatrix pairs, using probabilities analogous to \eqref{cr_distr}. Crucially, sketching introduces linear dependencies across the coordinates, which this method exploits to lower the recovery threshold. This shows how structured sampling and coding tools that utilize dependencies can jointly improve the efficiency and robustness of ACC.

This scheme provides a perspective in which exploiting structure in the data itself can further reduce redundancy, extending the benefits of sketching beyond purely random constructions.\\

\noindent\underline{\textbf{Summary of Section~\ref{ACMM_sec}}}: Across the ACMM schemes we presented, a unifying theme is the use of sampling and coding to efficiently approximate matrix products, where sketching reduces the dimensionality of the computation while redundancy mitigates the impact of stragglers. This is particularly well-suited for large-scale linear algebra workloads, where matrix multiplication is a core bottleneck and exact computation is often computationally prohibitive. The methods discussed reflect different trade-offs between approximation accuracy, recovery threshold, and computational and communication complexity.

\section{Conclusion}

This article surveyed key polynomial-based CC schemes for GC and CMM, as well as \textit{approximate} schemes that draw on ideas and techniques from sketching. At a high level, CC and sketching represent two complementary approaches to scalability: while CC introduces \textit{redundancy} to ensure erasure-tolerance, sketching constructs \textit{compressed} representations of data and matrices to reduce computational and communication costs. \textit{Approximate} CC bridges these paradigms by relaxing \textit{exact} recovery requirements, enabling efficient and scalable solutions that balance accuracy with system efficiency. After motivating both areas and presenting \textit{exact} CC schemes, we outlined \textit{approximate} schemes that combine ideas from both fields, showcasing their synergy and potential for scalable and resilient computation.

Rather than individual constructions, a central takeaway from this survey is the unifying role of redundancy and compression in enabling scalable distributed computations. Across the methods discussed, CC and sketching offer complementary mechanisms for addressing system bottlenecks, with approximate schemes bridging the two by relaxing exact recovery in favor of efficiency. Understanding these methods through this common lens helps clarify their respective strengths, limitations, and applicability across a range of large-scale optimization and linear algebra tasks.\\

\noindent{\textbf{Neophytos Charalambides}} is a postdoctoral fellow at the CSE department of UCSD, and received his PhD from the University of Michigan.

\noindent{\textbf{Arya Mazumdar}} is a professor at UCSD in the School of Computing, Information and Data Sciences. His research interests include mathematical statistics, information theory, learning, and coding theory.

\bibliographystyle{IEEEtran}
\bibliography{refs}


\end{document}